\documentclass[aps,pre,amsmath,amssymb,superscriptaddress,twocolumn]{revtex4-1}
\usepackage{amsmath}
\usepackage{amssymb}
\usepackage{wasysym}
\usepackage{pifont}
\usepackage{graphicx,overpic}
\usepackage{mathrsfs}
\usepackage{nicefrac} % Added for nicer fractions as arguments of functions, -Anshuman Dubey
\usepackage[normalem]{ulem}
\usepackage{hyperref}

\newcommand{\ed}{\ensuremath{\text{d}}}

\usepackage{xcolor}

\begin{document}

\title{Fractal dimensions and trajectory crossings in correlated random walks}
\author{A. Dubey}
\author{J. Meibohm}
\author{K. Gustavsson}
\author{B. Mehlig}
\affiliation{Department of Physics, University of Gothenburg, SE-41296 Gothenburg, Sweden}
\date{\today}

\begin{abstract}

We study spatial clustering in a discrete, one-dimensional, stochastic, toy model of heavy particles in turbulence and calculate the spectrum of multifractal dimensions $D_q$ as functions of 
a dimensionless parameter, $\alpha$, that plays the role of an inertia parameter. Using the fact that it suffices to consider the linearized dynamics of the model at small separations, we find that $D_q =D_2/(q-1)$
 for $q=2,3,\ldots$. The correlation dimension $D_2$ turns out to be a non-analytic function of the inertia parameter in this model. We calculate $D_2$ for small $\alpha$ up to the next-to-leading order in the non-analytic term.
\end{abstract}

\maketitle
\section{Introduction}

Heavy particles in turbulent flows occur frequently in nature. Examples are  small rain
droplets in turbulent rain clouds \cite{Dev12b}, microscopic sand grains in the turbulent gas surrounding growing stars \cite{Joh14, Wil08}, and microscopic  plankton in ocean
turbulence \cite{Rei03,Gua12, Gus16b, Cen13}. These turbulent aerosols show strong inhomogeneities in the spatial distribution of particles, in particular at small spatial scales \cite{Gus16}.
Such small-scale spatial clustering was observed in experiments \cite{War09, Dev12b}, and in direct numerical simulations \cite{Saw12b,Bec11}.

Heavy particles may detach from the turbulent flow, so that their trajectories can cross. Tracer particles that are constrained to follow a velocity field, on the other hand,  cannot cross paths, since any velocity field must be single-valued.
The crossing of trajectories, therefore, is an inertial effect where particle phase-space manifolds fold back upon themselves, causing multi-valued particle velocities that can give rise to large collision velocities \cite{Fal02,Sun97,Wil06,Gus11b,Gus13c,Vos13}.
The loci in space that delineate the multi-valued regions are referred to as \lq caustics\rq{}~\cite{Wil03,Wil05,Cri92}.

It is a challenge to describe turbulent aerosols from first principles because its analysis must take into account the underlying turbulence, a non-linear, out-of-equilibrium problem, with an infinite number of strongly coupled degrees of freedom \cite{Obe02}. Instead, a statistical approach to model particles in turbulence has been developed \cite{Fal01,Gus16}. In such statistical models one replaces the deterministic fluid velocity field by a smooth random function with prescribed statistics. In particular, statistical models of particles in turbulence have been useful in the study of small-scale clustering \cite{Gus11a,Bal01,Chu05,Dun05}
caustic formation \cite{Wil06,Gus13a}, and have significantly advanced our understanding of heavy-particle dynamics in turbulence.

However, two important issues remain unresolved. First, inertial particles in turbulence have been numerically shown to cluster on multifractal sets \cite{Bec11}, characterized by their multifractal dimensions $D_q$. The multifractal dimensions $D_q$ measure the degree of inhomogeneity in the distribution of the particles as power-laws to the $q-$th mass moments \cite{Man74,Hal86}. While the correlation dimension $D_2$ and the Lyapunov dimension (related to $D_1$) have been studied in quite some detail \cite{Gus16}, little is known about general multifractal dimensions $D_q$. They have been calculated in the case of tracer particles in compressible flows \cite{Bec04}. This study, however, excludes trajectory crossings, since the velocities of tracer particles are single-valued.  Second, the dependence of the correlation dimension $D_2$ on the inertia parameter (the Stokes number St) is not well understood. The correlation dimension $D_2$ shows a minimum as a function of St, an effect that is not captured by perturbation theory \cite{Wil10b,Gus15}. Recently, it was argued that the formation of caustics could be the reason for the failure of perturbation theory in $D_2$ \cite{Jan17}. 

In this paper, we analytically calculate the spectrum of multifractal dimensions $D_q$ and investigate the effect of the rate of trajectory crossings $J$ on the multifractal dimensions $D_q$. We consider a statistical toy model to study the clustering of particles suspended in a turbulent flow. The model is a one-dimensional, discrete-time random walk model \cite{Deu84,Deu85,Wil12a}, which describes the discrete dynamics of an ensemble of random walks immersed in a flow field. The flow is taken to be a smooth, random velocity field to model turbulence in the dissipative range. Due to the flow field, random walks that are spatially close to each other are correlated and may travel together for some time. The model includes effects that are similar to particle inertia and the turbulent flow, other effects such as particle size and particle-particle interactions are disregarded.
The dimensionless number $\alpha$ plays the role of an inertia parameter (the Stokes number St for heavy particles in turbulence). It is defined as the ratio of the mean-squared displacement and the correlation length.  The long-time distribution of random walks in this model exhibits a statistical steady state with multifractal clustering. This behavior is similar to that of heavy inertial particles in incompressible turbulence, where the correlated displacement of nearby particles results in small-scale spatial clustering.

The motivation for considering the one-dimensional, discrete-time random walk model as a playground is two-fold. Firstly, this model can be seen as a discretization of an over-damped, continuous, one-dimensional model of particles in turbulence. Secondly, we can analytically compute observables like $D_q$ and $J$, which is not possible in the continuous-time models in two and three spatial dimensions. This analytical control allows us to find exponentially small non-analytic contributions which could give insights into the physical phenomenon affecting clustering.

We find that the multifractal dimension spectrum $D_q$ is related to the way in which particle trajectories cross and derive a relation between the multifractal dimensions and the correlation dimension, $D_q = D_2/(q-1)$ for $q=2,3,\ldots$. The same relation holds for deterministic hyperbolic systems \cite{Gra83a,Gra83b}. For small $\alpha$, we use an implicit equation for the correlation dimension $D_2$ to derive a non-perturbative, asymptotic expansion of $D_2$ for small $\alpha$.

We note that the multifractal dimensions are defined in the mathematical limit of vanishing particle separations. This is an unrealistic assumption for physical systems where the finite particle size sets a lower limit on smallest relevant length scale of the system. Therefore to match with experiment one must be able to describe clustering at finite separations between particles, and non-divergent average densities.

The paper is organized as follows: In Section \ref{sec:model} we motivate the random-walk model from the continuous one-dimensional stochastic model of particles in turbulence and discuss the details of the random-walk model. Next, in Section \ref{sec:frac} we discuss the multifractal dimensions, and in Section \ref{sec:cordim} we present the non-perturbative expansion of the correlation dimension $D_2$ in the limit $\alpha \to 0$. In Section \ref{sec:crossings} we derive the rate of trajectory crossings in the linearized, as well as in the full non-linear model, and compare with results obtained from simulations. In Section \ref{sec:relation} we use trajectory crossings to derive a relation between the multifractal dimensions $D_q$ and the correlation dimension $D_2$. Section \ref{sec:conclusions} contains conclusions and discussions of the presented work. Technical details on finite-time Lyapunov exponents and the Mellin-Barnes transform are discussed in the appendices.

\section{Model}\label{sec:model}
We start with the equation of motion for an inertial particle in a one-dimensional continuous random flow \cite{Wil03},
\begin{align}
\dot{x} = v \quad\text{and}\quad \dot{v} = \gamma \left[u(x(t),t) - v\right],
\label{eq:xvStokes}
\end{align}
where $x$ and $v$ are the particle position and velocity respectively, and $u (x,t)$ is the fluid velocity at position $x$ and time $t$. Taking the overdamped limit $\gamma \to \infty$ reduces the two equations (\ref{eq:xvStokes}) to a single equation,
\begin{align} \label{eq:overdamped}
	\dot x(t) \sim u(x(t),t)\,,
\end{align}
the dynamics of tracer particles. Eq.~\eqref{eq:overdamped} shows that the continuous overdamped model is non-inertial and has no trajectory crossings since the particle follows the single-valued fluid field $u(x(t),t)$. However, discretizing this model at a non-infinitesimal time step $\Delta t$ reintroduces inertial effects. We obtain
\begin{align}
 x(t + \Delta t) = x(t) +  u(x(t),t)\Delta t.
\end{align}
Fixing $\Delta t = 1$, we end up with the discrete, iterative dynamics
\begin{equation}\label{eq:eom}
	x_{n+1} = x_{n} + f_n(x_n),
\end{equation}
where $x_n \equiv x(t_n)$ and $f_n(x_{t}) \equiv u(x(t_n),t_n)$. In order to model the spatial smoothness and the dynamics of the flow field $u(x(t),t)$ we take $f_n(x)$ to be a Gaussian random function with zero mean and correlation function
\begin{equation}\label{eq:corr}
	\langle f_m(x) f_n(0) \rangle = \delta_{mn} \sigma^2 \exp\left(-\frac{x^2}{2 \eta^2}\right)
\end{equation}
which defines the mean-squared displacement $\sigma$, and correlation length $\eta$. The brackets $\langle\ldots\rangle$ denote an average taken over a large ensemble of walkers with different initial conditions. The system size $L$ introduces an additional length scale in the system. We impose periodic boundary conditions with period $L$ on the equations of motion. The equation of motion can be dedimensionalized with the correlation length $\eta$ by changing coordinates according to $x\to \eta x$ and $f_n\to \eta f_n$. We find that the model depends on the dimensionless parameters
\begin{equation}
	\alpha \equiv \frac{\sigma}{\eta}\,, \quad \text{and} \quad l \equiv \frac{L}{\eta}\,.
\end{equation}
The multifractal steady-state distribution of particle positions is obtained by iterating an initial density of $N$ walkers.
$\{x^{(k)}_0, \ k=1,\dots,N\}$  a large number of times $n\gg1$ according to the dynamics \eqref{eq:eom}. 

\section{Multifractal dimensions}\label{sec:frac}
The fractal dimension spectrum $D_q$, where $q \in \mathbb{R}$, quantifies the nature of singularities of the spatial distribution $P(x_n)$ of the set of walkers $S_n$ and describes the inhomogeneity of the fractal \cite{Gra83a,Gra83b}. For instance, a fractal set with a homogeneous distribution of points has for all $q,q^\prime$,  $D_q = D_{q^\prime}$. However, in general $D_q \geq D_{q^\prime}$ if  $q < q^\prime$ \cite{Har01,Ott02}. The fractal dimensions are defined by the scaling relation
\begin{equation}\label{eq:Dq}
	\langle m_{x,\varepsilon}^{q-1}\rangle \sim \varepsilon^{(q-1)D_q}
\end{equation}
in the limit of $ \varepsilon\to 0$, where $m_{x,\varepsilon}$ is the number of walkers in an $\varepsilon$-interval around a reference walker located at $x$, and $\langle \cdots\rangle$ is an average obtained by using all walkers as reference walkers. For integer values of $q$ larger than one, $q=2,3,\ldots$, $D_q$ may alternatively be defined, and more efficiently calculated from simulations, by following the positions $\{x^{(k)}_n, \ k=1,\dots,q\}$ of $q$ walkers, given that $n\gg1$. We have \cite{Har01,Bec04}
\begin{align}\label{eq:Yqdef}
	\langle m_{x,\varepsilon}^{q-1} \rangle = P ( Y_n^{(q)} \leq \varepsilon)\,,
\end{align}
where the $Y^{(q)}_n$ are defined as
\begin{align} \label{eq:Yqdef_1}
	Y^{(q)}_n = \max_{1\leq i,j\leq q} \{ |x_n^{(i)}-x_n^{(j)}|\}\,.
\end{align}
In the limit of small separations it follows from Eq.~\eqref{eq:Dq} that
\begin{align}\label{eq:2nddefDq}
	P( Y^{(q)}_n \leq \varepsilon) \sim \varepsilon^{(q-1) D_q}\,,\qquad\varepsilon \ll 1\,.
\end{align}
In Section \ref{sec:relation} we use Eq.~\eqref{eq:2nddefDq} to find a relation between the correlation dimension $D_2$ and the multifractal dimensions $D_q$.  
\section{Correlation dimension} \label{sec:cordim}
The (fractal) correlation dimension $D_2$ is of particular physical importance. As Eq.~\eqref{eq:2nddefDq} suggests, $D_2$ measures the power-law singularity of the probability density of separations, $P(|\delta x_n|=\varepsilon)$, at small separations \cite{Gra83,Gra83a}. More precisely, we obtain by using Eq.~\eqref{eq:2nddefDq} the relation
\begin{align}\label{eq:dist}
	P(|\delta x_n| = \varepsilon) = \frac{\ed}{\ed \varepsilon}P(|\delta x_n| \leq \varepsilon) \sim \varepsilon^{D_2-1}\,.
\end{align}
The authors of Ref.~\cite{Wil12a} calculated $D_2$ for the present model, using a a short-time approximation for the Liouville operator. They derived the implicit formula
\begin{align}\label{eq:D2}
	\Gamma \left[\frac{-D_2+1}{2}\right]{_1F_1}\left[\frac{D_2}{2};\frac{1}{2};-\frac{1}{2 \alpha ^2}\right] = \pi^{\frac{1}{2}}(\sqrt{2}\alpha)^{D_2}\,.
\end{align}
It was shown that this relation admits a solution with $0<D_2<1$ for all $\alpha>\alpha_{\rm c}\approx1.56$, which is in line with the requirement that $P(|\delta x_n|)$ be normalizable.

\begin{figure}[t]
\includegraphics{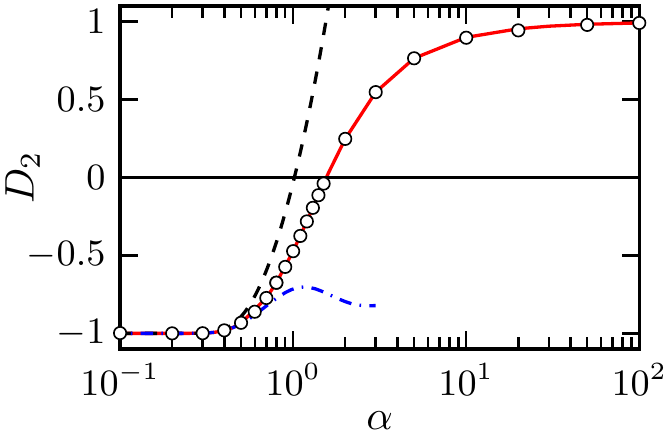}
\caption{Correlation dimension as a function of $\alpha$. Shown is the numerical solution of Eq.~(\ref{eq:D2}), solid red line, the leading-order of Eq.~(\ref{eq:d2_2}), dashed line, the resummation (see Appendix \ref{appendix:d2}) of the asymptotic approximation to $D_2$, Eq.~(\ref{eq:d2_1}), dash-dotted blue line, and results of numerical simulations (symbols).}\label{fig:d2}
\end{figure}
In turn, for $\alpha < \alpha_{\rm c}$, Eq.~\eqref{eq:D2} still has a non-trivial solution but with negative correlation dimension $D_2 < 0$. Fig.~\ref{fig:d2} shows $D_2(\alpha)$ as predicted by the relation \eqref{eq:D2} as the red solid line. A negative correlation dimension seemingly results in a non-normalizable density $P(|\delta x_n| = \varepsilon)$ due to the divergence at small separations according to Eq.~\eqref{eq:dist}.

As discussed in Refs.~\cite{Pik92,Wil15,Jan17} one can make sense of the correlation dimension also for $D_2 < 0$ by regularizing the dynamics at small separations. This is done by adding a weak additional random \lq{}noise\rq{} to Eq.~\eqref{eq:eom} according to
\begin{equation}\label{eq:noise}
	x_{n+1} = x_n+ f_n(x^{(1)}_n)+\xi_n\,.
\end{equation}
Here, $\xi_n$ are independent identically distributed Gaussian random variables with zero mean and variance $\langle \xi_n^2\rangle = \kappa^2$. As the noise $\xi_n$ is purely auxiliary, we choose $\kappa$ small, in particular, $\kappa \ll \alpha$. Therefore, $\xi_n$ cuts off the power law in \eqref{eq:dist} at small $\varepsilon$, which results in a uniform distribution at scales up to $\varepsilon \approx \kappa$. For $\kappa\ll\varepsilon\ll1$, on the other hand, the distribution of separations now follows the power law \eqref{eq:dist} with negative correlation dimension $D_2<0$ \cite{Wil15}. A negative $D_2$ is not in contradiction with $P(|\delta x|=\varepsilon)$ being normalizable because of the small-scale cut-off at $\varepsilon \approx \kappa$. A noise term similar to the one in Eq.~\eqref{eq:noise} has been shown to arise naturally in turbulent suspensions of heavy particles of different sizes \cite{Chu05,Bec05,Jan17}. The white symbols in Fig.~\ref{fig:d2} are results of numerical simulations for $D_2$ obtained by measuring the scaling of $P(|\delta x_n|=\varepsilon)$ for $\kappa\ll\varepsilon\ll1$, using the regularized dynamics \eqref{eq:noise}. We observe excellent agreement with the prediction provided by Eq.~\eqref{eq:D2} for both positive and negative $D_2$.

The numerics and the simulations shown in Fig.~\ref{fig:d2} suggest that $D_2\sim-1$ for $\alpha\ll1$. Corrections to this relation are found by making the ansatz $D_2\sim-1+\beta(\alpha)$ with $\beta(\alpha)\ll\alpha\ll1$ in Eq.~\eqref{eq:D2} and solving for $\beta(\alpha)$. We find the asymptotic expansion
\begin{align}\label{eq:d2_1}
	&D_2 \sim  -1 + \!\frac{{\rm e}^{-1/(2\alpha^2)}}{\sqrt{2\pi}}(4\alpha -14 \alpha^3 \!+63\alpha^5\!-\frac{905}{2} \alpha^7\!+\dots) \nonumber\\
	    +& \frac{{\rm e}^{-1/\alpha^2}}{2\pi}[16 \gamma  \alpha ^2\!-4 (3+28 \gamma ) \alpha ^4\!+(\frac{170}{3}+700 \gamma) \alpha ^6\!-\ldots]	\nonumber\\
	    +&\frac{{\rm e}^{-1/\alpha^2}}{2\pi}\log(\alpha^2/2)[ 	-8 \alpha ^2\!+56 \alpha ^4\!	-350 \alpha ^6\! +\ldots],
\end{align}
where $\gamma$ denotes the Euler-Mascheroni constant.  The details of the calculation that lead to Eq.~\eqref{eq:d2_1} can be found in Appendix \ref{appendix:d2}. The asymptotic expansion for $D_2$ in Eq.~\eqref{eq:d2_1} is shown in Fig.~\ref{fig:d2} as the blue dash-dotted line, and is seen to be an excellent approximation up to $\alpha \approx 0.4$. The subleading asymptotic terms of the order $\exp(-1/\alpha^2)$ and logarithmic contributions in Eq.~\eqref{eq:d2_1} have been obtained by using the Mellin-Barnes technique, which is described in more detail in Appendix \ref{sec:BMT}. 

Previous calculations of the correlation dimension performed using perturbation theory did not capture the non-analytic terms of the form $\exp(-1/{(2\alpha^2)})$. The leading order non-analytic contributions for $D_2 $ in the continuous one-dimensional model were obtained in \cite{Jan17}. In Eq.~\eqref{eq:d2_1} we have calculated the corresponding non-analytic contributions to next-to-leading order for the present model. Series of the form in Eq.~\eqref{eq:d2_1} occur in quantum mechanics and quantum-field theory, and are referred to as \lq trans series\rq{}  \cite{Zin04a, Zin04b, Dor14, Dun14}. 
\section{Crossing trajectories} \label{sec:crossings}

\begin{figure}[t]
	\includegraphics{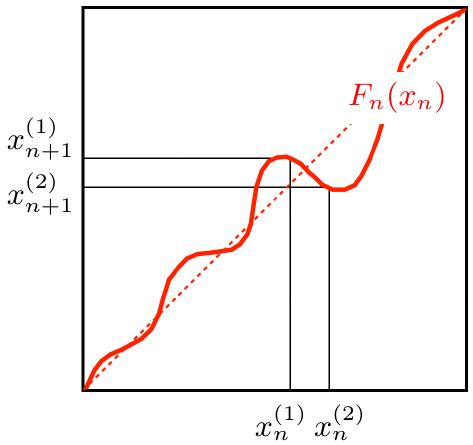}
	\caption{An illustration of crossing trajectories: $x_n^{(1)} < x_n^{(2)}$ is mapped to $x_{n+1}^{(1)} > x_{n+1}^{(2)}$. For close-by particles this occurs where $F_n' < 0$. The thick solid red line shows a realization of the random function $F_n(x_n)$. The dashed line shows $x_{n+1}=x_n$.}
\label{fig:cross}
\end{figure}
The iterative dynamics defined by Eq.~\eqref{eq:eom} allows for trajectory crossings of nearby random walkers. Crossings occur when the random map that generates the iterations,
\begin{align}\label{eq:F}
	F_n(x) \equiv x+ f_n(x),
\end{align}
has realizations that are not one-to-one. Since the realizations $F_n$ are smooth, a necessary condition for $F_n$ to be one-to-one is that its derivative is positive everywhere, $F_n'>0$. In turn, trajectories of nearby walkers may cross if there are finite regions in $x$ for which this derivative is not positive, $F_n'(x)\leq0$. Fig.~\ref{fig:cross} schematically depicts a realization $F_n(x)$ that is multivalued and two walkers whose trajectories cross.

The crossing rate $J$ for two infinitesimally close walkers can be obtained by linearizing the dynamics of separations of two particles. The separation $\delta x_n= x_n^{(1)}-x_n^{(2)}$ of two close-by walkers obeys the asymptotic dynamics
\begin{align}\label{eq:lindyn}
	\delta x_{n+1} \sim (1+A_n)\,\delta x_{n},	\qquad |\delta x_n|\ll1\,,
\end{align}
where $A_n$ are identically distributed Gaussian random variables with zero mean and variance $\langle A_n^2\rangle = \alpha^2$, see Appendix \ref{sec:ftle} for a related discussion of the calculation of Lyapunov exponents from the linearized dynamics. Crossings occur when the separation $\delta x_n$ changes sign between two subsequent time steps of \eqref{eq:lindyn}. Thus a sufficient condition for a crossing in the linearized dynamics is that $A_n<-1$. At each time step, the probability $P(A_n<-1)$ and, hence, the rate of crossing $J$ at small separation is given by
\begin{align}\label{eq:J}
J	&\sim P(A_n<-1)=\int_{-\infty}^{-1}\frac{{\rm d}A_n}{\sqrt{2\pi\alpha^2}}\, {\rm e}^{-A_n^2/(2\alpha^2)},	\nonumber \\
	&= \frac{1}{2}{\rm erfc}(1/\sqrt{2\alpha^2}),	\qquad |\delta x_n|\ll1,
\end{align}
where $\text{erfc}$ is the complementary error function.
For small $\alpha$, $J$ is exhibits an exponential activation according to
\begin{equation}\label{eq:J_asym}
J \sim \frac{\alpha \, {\rm e}^{-1/(2\alpha^2)}}{\sqrt{2\pi}},	\qquad \alpha\ll1\,.
\end{equation}

The rate of trajectory crossing $J$ is the analogue to the rate of caustic formation $\mathcal{J}$ in the corresponding one-dimensional continuous model for inertial particles in turbulence \cite{Gus13a,Gus16}. Interestingly, $\mathcal{J}$ shows a similar activation in the weak-inertia limit. Furthermore, using the leading-order term in Eq.~\eqref{eq:d2_1} and Eq.~\eqref{eq:J_asym} we find
\begin{equation}\label{eq:d2_2}
	D_2 \sim  -1 + 4J\,, \qquad \alpha\ll1\,.
\end{equation}
In the regime $\alpha \ll 1$ the Lyapunov exponent is negative, $\lambda < 0$ so a pair of trajectories converge towards each other. If the rate of trajectory crossings was identically $0$, the trajectories would eventually fuse together giving $D_2=-1$. A non-zero value of $J$ leads of the trajectories oscillating around each other and causing a spread in the distribution of particles, thereby reducing clustering i.e. forcing $D_2>-1$. A similar relation between $D_2$ and $\mathcal{J}$ has recently been found in the corresponding continuous white-noise model, namely $D_2\sim-1+2\mathcal{J}$ \cite{Jan17}.
\begin{figure}
	\includegraphics{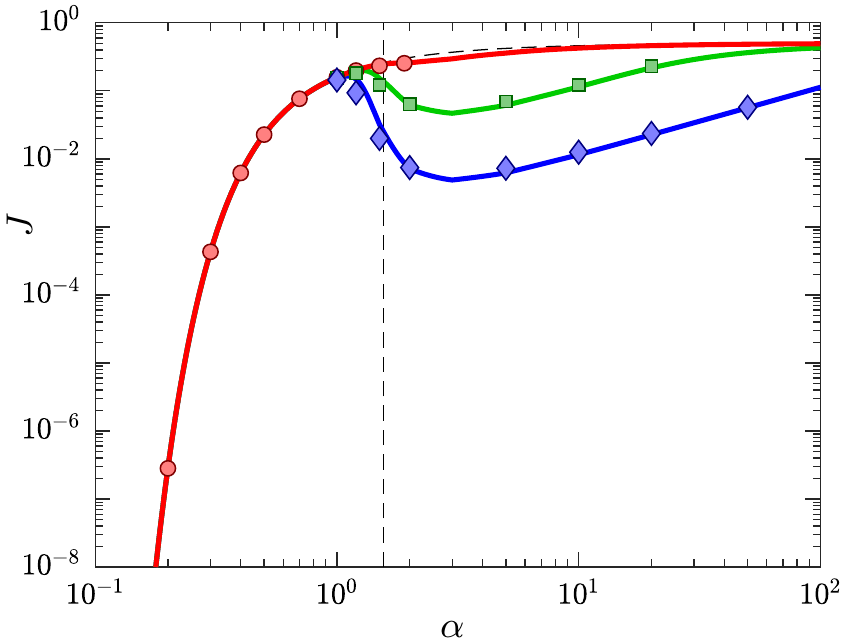}
	\caption{Probability of trajectory crossing $J$ against $\alpha$. The black dashed line shows the theoretical prediction from the linearized model, Eq.~\eqref{eq:J}. The colored lines show the rate of crossings obtained from the non-linear model theoretically, Eq.~\eqref{eq:J_full}, and from simulations with $l = 10$ (red circles), $l = 10^2$ (green boxes), $l =10^3$ (blue diamonds). The vertical black line shows $\alpha_{\rm c}=1.56$}\label{fig:J}
\end{figure}

The expression for $J$ given in \eqref{eq:J} is a good approximation for the rate of trajectory crossings in the full (non-linear) model as long as most crossings occur between nearby trajectories. That is the case if $\alpha<\alpha_\text{c}$, when the Lyapunov exponent of the system is negative and trajectories spend most of their time close together (see Appendix \ref{sec:ftle}) regardless of the size of $l$. For $\alpha>\alpha_\text{c}$, in turn, trajectories spend most of their time far apart so that crossings are more likely to occur at larger separations. Moreover the likelihood of two trajectories traveling away from each other increases as $l$ increases. This leads to an increase in deviations from the linearized rate of crossings as $l$ increases. Crossings at larger separation are not described by the linearized model, Eq.~\eqref{eq:lindyn}. Fig.~\ref{fig:J} shows the probability of trajectory crossing in the linearized dynamics $J$ (black dashed line) and numerical simulations of the full model (markers) as functions of $\alpha$ for different values of $l$. The data suggests that Eq.~\eqref{eq:J} is an excellent approximation to the exact probability of trajectory crossing up to values $\alpha\approx\alpha_{\rm c}$.

The rate of trajectory crossings in the full non-linear model can be obtained as follows. The equation of motion for the separation between two particles $\delta x_n$ is
\begin{align}\label{eq:sepeq}
	\delta x_{n+1} &= \delta x_n + \delta f_n,
\end{align}
where $\delta f_n := f_n(x_n^{(1)}) - f_n(x_n^{(2)})$. The probability of a trajectory crossing is given by the probability of the separation changing sign between two subsequent time steps. That means
\begin{multline}
 J = P(\delta x_{n+1} < 0 ; \delta x_n >0 ) \\
+P( \delta x_{n+1} > 0 ; \delta x_n <0 ),
\label{eq:JFiniteContributions}
 \end{multline}
where $P(A;B)$ is the joint probability of events $A$ and $B$. Using particle-interchange symmetry, Eq.~\eqref{eq:sepeq}, and factorization of the joint probability due to the independence of $\delta f_n$ and $\delta x_n$,  we can write Eq. (\ref{eq:JFiniteContributions}) as
 \begin{align}
 	J =2 \int_0^\infty \text{d}\varepsilon \ P( \delta f_n <-\varepsilon) P(\delta x_n = \varepsilon).
 \end{align}
Because $\delta f_n$ is a sum of two Gaussian random functions, it is itself a Gaussian random function, with zero mean and variance
\begin{align}
	\langle(\delta f_n)^2\rangle = 2\alpha^2(1-e^{-{\delta x_n^2}/{2}})\equiv v(\alpha,\delta x_n),
\end{align}
so that $P( \delta f_n <-\varepsilon) = \text{erfc}(\varepsilon/\sqrt{2 v(\alpha,\varepsilon)})/2$. Using integration by parts, this gives 
\begin{align}\label{eq:J_full}
	J  = - \frac12\int_0^\infty\!\!\! \ed \varepsilon P(|\delta x_n| \leq \varepsilon) \frac{\ed}{\ed \varepsilon} {\rm erfc}\left( \frac{\varepsilon}{\sqrt{2 v(\alpha,\varepsilon)}}\right).
\end{align}
Note first that for $\varepsilon \ll 1$ the term $\text{erfc}(\varepsilon/\sqrt{2 v(\alpha,\varepsilon)})$ has the asymptotic form
\begin{align}
	\text{erfc}(\varepsilon/\sqrt{2 v(\alpha,\varepsilon)})\sim \text{erfc}(1/\sqrt{2\alpha^2}).
\end{align}
Using this, we infer from Eq.~\eqref{eq:J_full} that $J= \text{erfc}(1/\sqrt{2\alpha^2})/2$ as in the linear model (see Eq.~\eqref{eq:J}) if $P(|\delta x_n| = \varepsilon)$ is concentrated at $\varepsilon=0$. That is the case when $\alpha<\alpha_\text{c}$, so that the rate of crossings in the full model reduces to the one obtained from the linearized model for $\alpha<\alpha_\text{c}$, as expected.

For $\alpha>\alpha_\text{c}$, $P(|\delta x_n| \leq \varepsilon)$ is a non-trivial function of $\varepsilon$ so that $J$ differs from the rate of trajectory crossings in the linearized model Eq.~\eqref{eq:J}. More precisely, we see that $J\leq \text{erfc}(1/\sqrt{2\alpha^2})/2$ for all $\alpha$ since $P(|\delta x_n|\leq \varepsilon)\leq 1$.

In order to evaluate $J$ for $\alpha>\alpha_\text{c}$ we need an expression $P(|\delta x_n| \leq \varepsilon)$ that is valid for  any value of $\varepsilon$. Eq.~\eqref{eq:dist}, however, is valid only at small $\varepsilon$. Including the small-scales noise and because we expect particles to be uncorrelated at larger $\varepsilon$, we make the following ansatz for the cumulative probability distribution:
\begin{align}\label{eq:comdist}
 P(|\delta x_n|\leq\varepsilon) =   \frac1l	\!\!	\times	\!\!
	\begin{cases}
		\varepsilon (x^*/x_0)^{1-D_2}\!\!\! &	\mbox{if } 0 <\varepsilon  \leq x_0,	\\
		\varepsilon^{D_2}{x^*}^{1-D_2}  & \mbox{if } x_0 <\varepsilon \leq x^*\!,	\\
		\varepsilon  	& \mbox{if } x^* <\varepsilon \leq l.	\\
	\end{cases}
\end{align}
Here $x_0$ is a small length scale related to the regularizing noise in Eq.~\eqref{eq:noise}, and $x^*$ is an arbitrary matching scale of order unity for the transition between the power law behavior of $P(|\delta x_n|\leq\varepsilon)$ and the large-scale uniform behavior. For $|\delta x_n| < x_0$ the dynamics is dominated by the noise term, and so the distribution is uniform. The solid lines in Fig.~\ref{fig:J} show results of numerical integration of Eq.~\eqref{eq:J_full} using Eq.~\eqref{eq:comdist} with $x^*=3$. We observe excellent agreement with simulations.

 \begin{figure}[t]
   \includegraphics{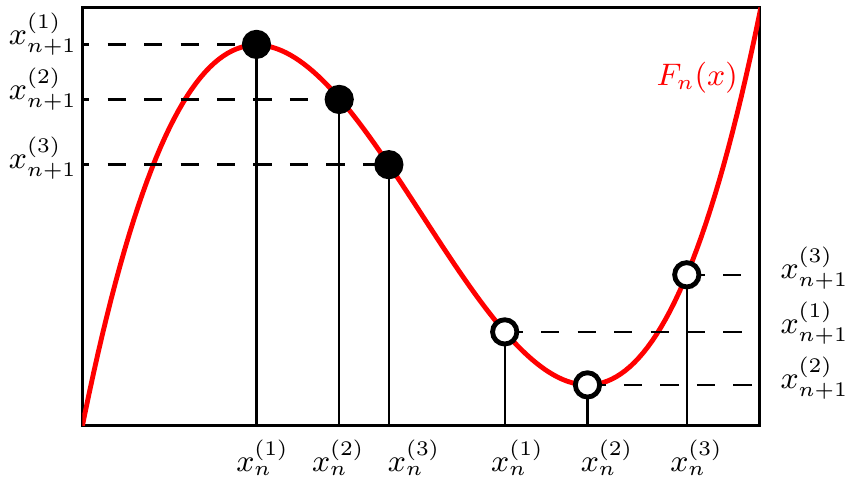}
\caption{\label{fig:c} Illustrates different types of crossings.
When the particles are in an approximately linear regime of the flow or close enough together (filled bullets on the left), then the identity of the
pair with maximal position distance does not change. When the particles
are further apart so that the non-linearity of $F_n$ becomes important,
the identity of the pair with maximum position distance may change (white bullets on the right).}
\end{figure}

In what follows, we denote the crossings that are governed by the linearized model Eq.~\eqref{eq:lindyn} \lq{}linear crossings\rq{}. Since we expect non-linear terms to play a role at larger $\alpha$, we call crossings that are not described by Eq.~\eqref{eq:lindyn} but only by the full (non-linear) model Eq.~\eqref{eq:eom} \lq{}non-linear crossings\rq{}, Fig. \ref{fig:c}.
\section{A relation between multifractal dimensions using trajectory crossings} \label{sec:relation}

Having discussed the multifractal dimensions and trajectory crossings in our model we now turn to the effects of trajectory crossings on clustering. We find that arguments about trajectory crossings lead to a relation between the multifractal dimensions $D_q$.

\begin{figure*}[t]
\includegraphics[clip,width=0.9\linewidth]{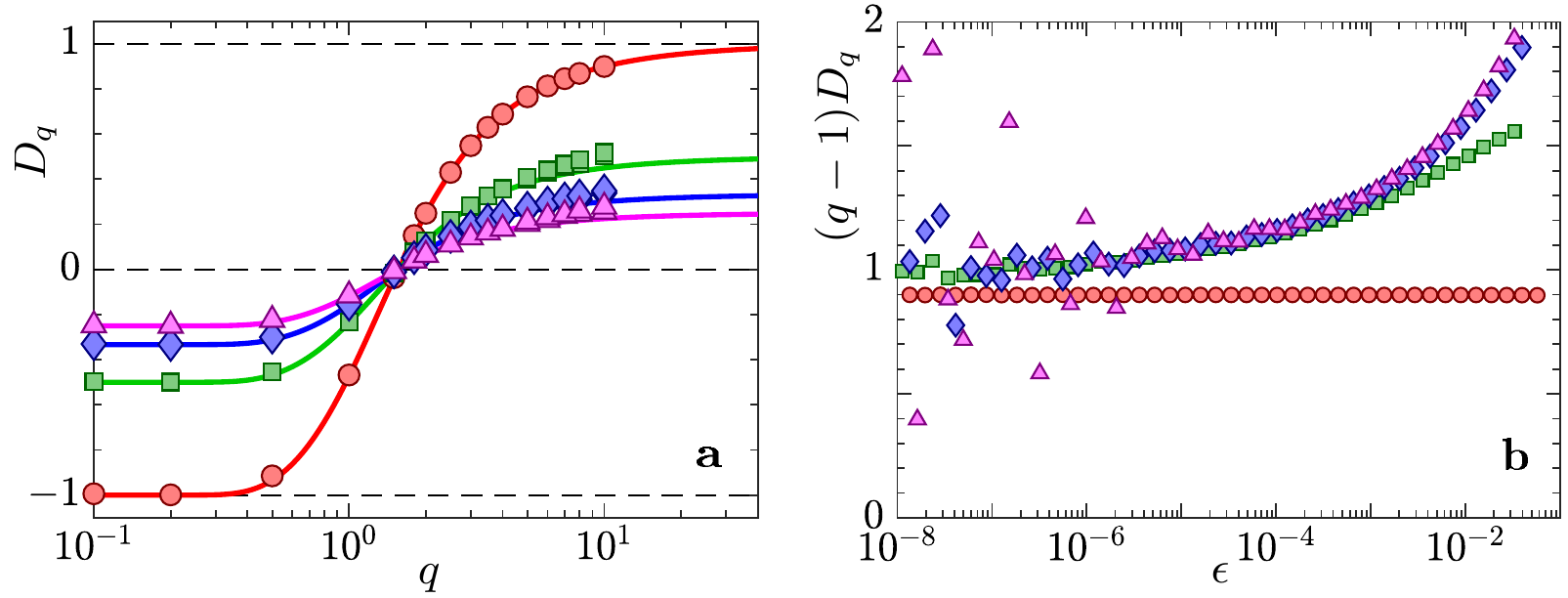}
\caption{Generalized fractal dimensions determined from numerical simulations using Eqs.~(\ref{eq:Dq}) and~(\ref{eq:Yqdef}). Panel \textbf{a} shows $D_q$ as functions of $\alpha$. The solid lines represent the theoretical predictions obtained using Eqs.~\eqref{eq:D2} and \eqref{eq:dqd2} and the symbols represent numerical simulations. Circles (red) correspond to $q=2$, squares (green) to $q=3$, diamonds (blue) to $q=4$, and triangles (pink) to $q=5$. Panel \textbf{b} shows numerical simulation data for the local scaling exponent $(q-1)D_q$ in the relation $P( Y \leq \varepsilon ) \sim \varepsilon^{(q-1) D_q}$ as a function of $\varepsilon$ for $\alpha=10$. Marker shapes corresponds to $q=2,3,4,5$ as in panel \textbf{a}. Horizontal dashed line shows $D_2=0.90$.}\label{fig:Dq}
\end{figure*}
First, consider how multifractal dimensions $D_q$ as defined in Sec.~\ref{sec:frac} are related for $q=2,3,\ldots$. As Eq.~\eqref{eq:J_asym} suggests the probability of trajectory crossings is exponentially small for $\alpha\ll1$. As a first approximation let us assume that there are no crossings for small $\alpha$. This approximation has interesting consequences for the $Y^{(q)}_n$ defined in Eq.~\eqref{eq:Yqdef_1}. Namely, if particle trajectories do not cross, the particle pair that has the largest separation at time step $n$ will have the largest separation also at any later time step, thus $Y^{(q)}_n=Y^{(2)}_{n}$, where $Y^{(2)}_{n}$ is obtained by considering only the two particles which initially had the largest separation. This means that we may follow the trajectories of the walker pair with the largest separation, without considering the other walkers. As a consequence we get for $Y^{(q)}_n$, $n\gg1$:
\begin{align} \label{eq:max2particles}
	P( Y^{(q)}_n \leq \varepsilon ) \propto P ( Y_n^{(2)} \leq \varepsilon ).
\end{align}
Now consider the possibility of linear crossings as defined in the end of Sec.~\ref{sec:crossings}. To that end, we name the walkers according to their order at step $n$, i.e. here and in the following,
\begin{align}
	x_n^{(1)}\leq x_n^{(2)} \dots \leq x_n^{(q)}\,.
\end{align}
It follows that at step $n$, $Y^{(q)}_n=|x_n^{(1)}-x_n^{(q)}|$. If the mutual separations of all walkers are small, in which case $Y^{(q)}_n\ll1$, crossings between the trajectories of $x^{(1)}$ and $x^{(q)}$ are almost always of the linear kind. A linear crossing of $q$ walkers at time step $n$ leaves the outermost walkers invariant, i.e., the crossing changes $x_n^{(1)}$ to $x_{n+1}^{(q)}$ and $x^{(q)}_n$ to $x^{(1)}_{n+1}$. Hence, $Y^{(q)}_{n+m}=Y^{(2)}_{n+m}$ for all $m$ if all crossings are linear, just as in the case of no crossings. It follows that Eq.~\eqref{eq:max2particles} holds also if $Y^{(q)}_n\ll1$ for all $n\gg1$. Using Eqs.~\eqref{eq:max2particles} and \eqref{eq:2nddefDq} gives the following relation between $D_2$ and $D_q$:
\begin{align} \label{eq:dqd2}
  D_q &= \frac{D_2}{q-1}.
\end{align}
Note that one can show that $D_q\geq 1/(1-q)$, see Appendix \ref{sec:ftle}. Using Eq.~(\ref{eq:dqd2}) it is trivial to generalize the relations \eqref{eq:d2_1} and \eqref{eq:d2_2} to all $D_q$ with $q=2,3,\ldots$. Note that the argument above holds true for any finite value of $\alpha$ because we may always take $\varepsilon$ in Eq.~\eqref{eq:max2particles} small enough so that $Y^{(q)}_n\ll1$ for all $n\gg1$. In practice this means, however, that as $\alpha$ becomes larger and the Lyapunov exponent $\lambda$ increases, it becomes increasingly hard to verify Eq.~\eqref{eq:dqd2} numerically. 
Fig.~\ref{fig:Dq}{\bfseries a} shows a comparison of $D_q$ in Eqs. (\ref{eq:dqd2}) and \eqref{eq:D2} to $D_q$ evaluated from the scaling exponent in Eq. (\ref{eq:2nddefDq}) using numerical simulations.
We observe good agreement within the limits of numerical accuracy if $\alpha$ is not too large or if $q=2$. 
For $q>2$ and $\alpha>\alpha_{\rm c}$ the convergence of the simulations to theory is slow and very small scales must be resolved, see Fig~\ref{fig:Dq}\textbf{b} and discussion below.

We remark that we have not considered non-linear crossings in our discussion. This kind of crossing in general does not leave the outermost walker pair invariant. In our simulations it was possible to observe non-linear crossings already at small but finite values of $\varepsilon$. In Fig~\ref{fig:Dq}\textbf{b} the scaling exponent $(q-1)D_q$ is measured numerically as a function of $\varepsilon$ for $\alpha>\alpha_{\rm c}$.
For $q=2$ the expected value $(q-1)D_q=D_2$ from Eq. (\ref{eq:dqd2}) is obtained for moderately large separations $\varepsilon$. 
For $q>2$, the simulation data converge slowly towards this value as $\varepsilon$ decreases and we observe a significant deviation from Eq.~\eqref{eq:dqd2} already at small but finite $\varepsilon$, which we attribute to the occurrence of non-linear crossings in the model.

Physical systems are typically equipped with a natural cutoff scale $\varepsilon_0$ that may be, for example, a finite walker size or a regime where small-scale diffusion dominates. We expect the physically relevant observables in realistic systems to be not the $\varepsilon \to 0$ scaling exponents $D_q$ but suitably defined, finite size counterparts $D_q(\varepsilon_0)$. A more detailed study of $D_q(\varepsilon_0)$ is left for future work.
\section{Conclusions} \label{sec:conclusions}
We studied clustering of heavy particles in turbulence by means of a simple one-dimensional discrete-time model. As the main result, we derived an intriguing relation between the multifractal dimensions $D_q$ and the correlation dimension $D_2$, $D_q=D_2/(q-1)$ and verified it for different values of $q$ by numerical simulations. A related expression for $D_q$ has been previously derived for hyperbolic systems without trajectory crossings in \cite{Bec04}. We show here that it is valid also for the present model, which is non-hyperbolic and allows for trajectory crossings. Furthermore, the derivation of Eq.~\eqref{eq:dqd2} via Eq.~\eqref{eq:2nddefDq} leads to the important insight that $(q-1)D_q = D_2$ holds true only if the rate of non-linear crossings is negligible compared to the rate of linear crossings. Mathematically, this is ensured because the multifractal dimensions $D_q$ are \textit{defined} in the limit of infinitesimal separations, $\delta x \to 0$, where all crossings are linear.

However,  to observe the relation $D_q = D_2/(q-1)$ in real systems one would need to have an large number of particles in a given volume. In contrast, systems of interest such as turbulent aerosols typically contain only a small number density of particles, of the order of a few of particles per Kolmogorov length cubed \cite{Sie15}. This leads to the conclusion that it would be interesting to study clustering for small average particle densities. Due to the natural small scale cut-off imposed by the particle size, a relevant quantity to calculate would be multifractal dimensions at non-zero separation, $D_q(\varepsilon)$, instead of the usual $D_q$ defined at infinitesimal separations.

Further, we analyzed the correlation dimension $D_2$ in the limit $\alpha\to0$. We found that naive perturbation theory fails because $D_2$ is non-analytic at $\alpha=0$. Our results indicate that the small-$\alpha$ expansion of $D_2$ is a trans-series of the general form
\begin{equation}\label{eq:trans}
D_2 = \sum_{k,l,m}  c_{klm} {\rm e}^{-k/(2\alpha^2)} \alpha^l \log^m (\alpha^2/2) \,.
\end{equation}
An intensively studied example of a trans-series is the quantum-mechanical energy spectrum of a particle in a  double well \cite{Ben69, Ben73}. There, the exponential contributions have a clear physical interpretation in terms of instantons, that is, collective excitations due to the presence of the degenerate potential minima. The corresponding power-series, in turn, are related to fluctuations around these instantons whereas logarithmic corrections are due to so called quasi-zero-modes \cite{Dun14}. In the present model, Eq.~\eqref{eq:d2_2} suggests that the multi-valuedness caused by crossing trajectories gives rise to similar instanton contributions. We expect that perturbation expansions for heavy-particle dynamics in turbulence have a similar structure, for expansions in the Stokes number, and also for related perturbation expansions in the white-noise limit \cite{Gus16,Jan17}. This would explain why the perturbation calculations of the correlation dimension in Refs.~\cite{Wil10b,Wil14} appear to miss important contributions. More generally, our results give novel insight into the mathematical structure that links fractal clustering with caustic formation in the dynamics of heavy particles in turbulence, since the singularities that make crossing of trajectories possible in the random-walk model correspond to caustic singularities in turbulent aerosols.

\appendix

\section{Finite-Time Lyapunov exponents}\label{sec:ftle}
The probability density function $P(\lambda_n)$ of finite-time Lyapunov exponents $\lambda_n$ characterizes the leading asymptotic behavior of particle pairs after a large number of time steps $n \gg 1$ of the dynamics \eqref{eq:eom}. This distribution is assumed to have the large deviation form \cite{Tou09}
\begin{align}
	P(\lambda_n) \approx {\rm e}^{-n I(\lambda_n)}\,,
\end{align}
where $I(\lambda_n)$ is called \lq{}rate function\rq{}. The infimum of $I(\lambda_n)$ determines the most likely value, $\lambda$, that $\lambda_n$ takes after $n \gg 1$ iterations. We call this value $\lambda$ such that $I(\lambda) = \inf_{\lambda_n} I(\lambda_n)$ the (ordinary) Lyapunov exponent. It is defined in the strict limit $n\to \infty$ according to
\begin{equation}\label{eq:lya}
\lambda \equiv \lim_{|\delta x_0|\to 0}\lim_{n\to\infty} \Big\langle \log \Big|\frac{\delta x_n}{\delta x_0}\Big|\Big\rangle\,,
\end{equation}
where $\delta x_n$ is the $n^\text{th}$ iteration of the initial separation $\delta x_0$. As shown in \cite{Wil12a}, $\lambda$ can be evaluated explicitly in the present model by considering the equation of motion for separations $\delta x_n = x^{(1)}_n - x^{(2)}_n$ of the momentary position of the two particles after $n$ iterations, $x^{(1)}_n$ and $x^{(2)}_n$, respectively. In this section, we extend the calculation in \cite{Wil12a} by providing an analysis of the distribution of finite-time Lyapunov exponents. We start out by using Eq.~\eqref{eq:eom} to derive an equation for $\delta x_n$ given by
\begin{align}\label{eq:eomdx}
	\delta x_{n+1} 	&= \delta x_{n} + f_n(x^{(1)}_n) - f_n(x^{(2)}_n)\,.
\end{align}
Linearizing the smooth function $f_n$ for small separations $\delta x_n\ll1$ we readily obtain
\begin{align}\label{eq:rec}
	\frac{\delta x_{n+1}}{\delta x_{n}} \sim 1+f_n'(x^{(1)}_n)\,.
\end{align}
Because $f_n$ and $f_m$ are uncorrelated for $n\neq m$, we can neglect the dependence of the right-hand side on $x^{(1)}_n$. This way, $f'_n(x^{(1)}_n)\equiv A$ simplifies to a single Gaussian random variable with $\langle A \rangle=0$ and variance $\langle A^2 \rangle = \alpha^2$. Using this, we solve the iteration in Eq.~\eqref{eq:rec} and get
\begin{align}
	\log \left|\frac{\delta x_n}{\delta x_0}\right| \sim \sum_{k=1}^{n} \log |1+A|\,.
\end{align}
By assuming ergodicity, we can replace the ensemble average over initial separations in Eq.~\eqref{eq:lya} by the sample mean of $\log |1+A|$. We find a simple equation for the finite-time Lyapunov exponent given by
\begin{equation}\label{eq:sam}
	\lambda_n = \frac{1}{n}\sum_{k=1}^n \log |1+A| \,, \qquad n\gg1\,.
\end{equation}
We use the Varadhan method \cite{Tou09} to obtain the rate function for the sample mean in Eq.~\eqref{eq:sam}. To this end, we first calculate the moment generating function of $\log |1+A|$ according to
\begin{align}
&\langle {\rm e}^{k \log |1+A|} \rangle = \int_{-\infty}^\infty\frac{{\rm d}A}{\sqrt{2\pi\alpha^2}}\, {\rm e}^{-A^2/(2\alpha^2)}\, |1+A|^k \nonumber\\
&=\pi^{-\frac{1}{2}}2^{\frac{k}{2}} \alpha ^{k} \Gamma \left[\frac{k+1}{2}\right] \, _1F_1\left[-\frac{k}{2};\frac{1}{2};-\frac{1}{2 \alpha ^2}\right]\,,
\end{align}
where $\Gamma$ is the gamma function and $_1F_1$ is the Kummer hypergeometric function. According to Varadhan's theorem, the cumulant generating function $\Lambda(k) = \log \langle {\rm e}^{k\log|1+A|}\rangle$ and the rate function $I(\lambda_n)$ are related by Legendre transform. We have
\begin{align} \label{legendt}
 I(\lambda_n) = \sup_{k} \{ \lambda_n k-\Lambda(k) \}.
\end{align}
Because $\Lambda(k)$ is a smooth function, we can replace the supremum in Eq.~\eqref{legendt} by the maximum and the relation $\lambda_n = \Lambda'(k)$ holds.

The Lyapunov exponent $\lambda$ given in Ref.~\cite{Wil12a} is recovered from the moment generating function:
\begin{align}\label{eq:lam}
	\lambda 	&= \frac{\ed}{\ed k}\bigg|_{k=0}\langle {\rm e}^{k \log|1+A|} \rangle \,, \nonumber\\
			&= \int_{-\infty}^\infty\frac{{\rm d}A}{\sqrt{2\pi\alpha^2}}\, {\rm e}^{-A^2/(2\alpha^2)}\, \log|1+A|\,.
\end{align}
Evaluation of the integral shows that $\lambda<0$ for small $\alpha$ and that it changes sign at $\alpha_{\rm c}\approx 1.56$ to become positive at larger $\alpha$. This transition was called \lq{}path-coalescence transition\rq{} \cite{Wil03,Wil12a}, because all paths coalesce in the limit of $n\to\infty$ for $\alpha<\alpha_{\rm c}$.
Below we derive a relation between the distribution of finite-time Lyapunov exponents $\lambda_n$ and the fractal dimension spectrum $D_q$ for our model. From arguments based on large-deviation theory, Pikovsky \cite{Pik92} derived an expression for the correlation dimension $D_2$ that applies directly to our model. In terms of the cumulant generating function $\Lambda(k)$ the author showed that
\begin{equation}\label{eq:c2}
	\Lambda(-D_2)=0\,.
\end{equation}
Using Eq.~\eqref{eq:dqd2} we can generalize this relation to $q=2,3,\ldots$ according to
\begin{equation}\label{eq:LambdaDq}
	\Lambda\big(-(q-1) D_q\big)=0\,.
\end{equation}
Condition \eqref{eq:LambdaDq} is equivalent to
\begin{align} \label{eq:c1}
	\min_{\lambda_n}\{\lambda_n (q-1)D_q + I(\lambda_n)\} = 0\,.
\end{align}
Related expressions were first derived for deterministic hyperbolic systems \cite{Gra83a,Gra83b}, and for particles advected
in compressible random velocity fields \cite{Bec04}. The discussion given here shows that the relations \eqref{eq:c2} and \eqref{eq:c1} also apply to the present system which includes trajectory crossings and is not deterministic and non-hyperbolic.

Eq.~\eqref{eq:LambdaDq} can alternatively be written as an integral fluctuation relation for the random quantity $(q-1)D_q \log|1+A|$ according to
\begin{align}\label{eq:FT}
	\langle {\rm e}^{-(q-1)D_q \log|1+A|}\rangle = 1\,.
\end{align}
Using Jensen's inequality and $\lambda=\langle \log|1+A| \rangle$ (see Eq.~\eqref{eq:lam}), it follows directly that
\begin{align}
	D_q \lambda \geq 0 \,,
\end{align}
for $q=2,3,\ldots$. This clearly shows that if $\lambda<0$, we must have $D_q<0$ and vice versa, in accordance with observation.

Furthermore, using Eq.~\eqref{eq:c1}, we now show that $D_q$ is bounded from below by $(1-q)^{-1}$, which can be observed in Fig.~\ref{fig:Dq}. Consider the \lq{}time-reversed\rq{} linearized dynamics (compare Eq.~\eqref{eq:lindyn})
\begin{align}\label{eq:linrev}
	\delta x_{n-1} \sim (1+A)^{-1} \delta x_n\,,
\end{align}
where $A$ is the Gaussian random variable defined above. For this time reversed process, the distribution of finite time Lyapunov exponents, Eq.~\eqref{eq:sam}, is identical to the \lq{}time-forward\rq{} case but with an overall minus sign. Eq.~\eqref{eq:c1} thus implies that the fractal dimension for the reversed process flips sign compared to the time-forward process. As all fractal dimensions, including that of the time-reversed process, are bounded above by one, this observation together with Eq.~\eqref{eq:dqd2}, leads to $-D_q(q-1)\leq1$. We thus find the lower bound
\begin{align}
	\frac{1}{1-q}\leq D_q \,,
\end{align}
for the time forward process.
\section{Non-analyticity of \texorpdfstring{$D_2$}{D2}} \label{appendix:d2}
We start from Eq.~(\ref{eq:c2}). A direct perturbation expansion of the correlation dimension
yields that $D_2=-1$ for $\alpha=0$, and that all other
perturbation coefficients vanish. The same happens in the one-dimensional
white-noise models for turbulent aerosols analyzed in
Refs.~\cite{Wil14,Wil15,Jan17}.
The main issue with local perturbation theory in this case is that if $D_2(\alpha)$ near $\alpha = 0$ is a non-analytic function, then a perturbative expansion would naturally fail.
One must use non-perturbative methods to analyze Eq.~(\ref{eq:c2}). To extract this non-analytic dependence we write
$D_2 = -1+\beta(\alpha)$. We assume that the $\beta$-term is small, insert this ansatz into Eq.~(\ref{eq:c2}), and expand
the condition (\ref{eq:c2}) in $\beta$:
\begin{align}
\label{eq:Iexpand}
	1 = I_0 - \beta I_1 + \beta^2 I_2 + \ldots{}\,.
\end{align}
The integrals $I_k$ are given by
\begin{equation} \label{eq:Ik}
I_k = \frac{1}{k!}\int_{-\infty}^\infty\frac{{\rm d}A}{\sqrt{2\pi\alpha^2}}\, {\rm e}^{-A^2/(2\alpha^2)}\, |1+A|\log^k|1+A|\,.
\end{equation}
Let us first consider the linear order in $\beta$.
Solving Eq.~(\ref{eq:Iexpand}) for $\beta$ we find to this order
\begin{align}
\label{eq:beta1}
\beta_1 &= \frac{I_0-1}{I_1}\,.
\end{align}
To compute the expansion in $\alpha$ we require the
asymptotics of $I_0$ and $I_1$. $I_0$ is given by the exact expression
\begin{align}
 I_0 = {\rm erf}\left( \frac{1}{\sqrt{2}\alpha}\right) + \frac{{\rm e}^{-1/(2\alpha^2)}}{\sqrt{2\pi}} 2\alpha,
\end{align}
and its series expansion reads:
\begin{align}
I_0 	= 1-\frac{{\rm e}^{-1/(2\alpha^2)}}{\sqrt{2\pi}} 2\alpha\sum_{k=1}^\infty (-1)^k (2k-1)!!\, \alpha^{2k}\,.
\end{align}
The expansion of $I_1$ is more difficult. We obtain the asymptotic expansion of $I_1$ up to the special function ${_1F_1}^{(1,0,0)}(a,b,z)$ (the superscript $(1,0,0)$ denotes a derivative in the first argument $a$, see main text)
\begin{multline}
 I_1(\alpha) \sim -\frac{1}{2} {\rm erf} \left[\frac{1}{\sqrt{2\alpha^2}}\right] \left\lbrace\gamma + \log\left(\frac{1}{2 \alpha^2}\right) \right\rbrace \\ +\frac{\alpha}{\sqrt{2 \pi}}  \exp\left[ -\frac{1}{2 \alpha^2}\right] \bigg\lbrace - \gamma - \log\left(\frac{1}{2 \alpha^2}\right)\\ + {_1F_1}^{(1,0,0)}\left(1,\frac{1}{2},\frac{1}{2 \alpha^2}\right)\bigg\rbrace,
\end{multline}
where $\gamma$ is the Euler--Mascheroni constant. We use the Mellin-Barnes integral representation of the Confluent Hypergeometric function $_1 F_1$ to obtain the asymptotics of ${_1F_1}^{(1,0,0)}$ to exponential accuracy, see Appendix \ref{sec:BMT}. We obtain
\begin{multline}\label{eq:I1_exp}
 	I_1(\alpha) 	= \sum_{k=0}^\infty \frac{(2k-1)!!}{2(k+1)} \alpha^{2k+2} + \frac{{\rm e}^{-1/(2\alpha^2)}}{\sqrt{2\pi}} \times \\
	 [2 (\gamma - 1)- \log(\frac{\alpha^2}{2})] \left(\sum_{k=1}^{\infty}(-1)^k (2k-1)!!\alpha^{2k+1}\right),
\end{multline}
for small positive values of $\alpha$. Expanding the integral Eq.~\eqref{eq:Ik} for $k=2$, we obtain the expansion for $I_2$ given by
\begin{multline}\label{eq:I2_exp}
	I_2(\alpha) = \frac{\alpha^2}{4} \sum _{k=0}^{\infty} \frac{(2 k+1)!! \left(1-H_{2 k}\right)}{ (k+1) \left(k+\frac{1}{2}\right)} \alpha^{2 k} \\+ \text{e}^{-1/(2\alpha^2)}(\ldots),
\end{multline}
where $H_{2k} = \sum_{n=1}^{2k} 1/n$ and $H_0=0$. The exponentially small corrections in Eq.~\eqref{eq:I2_exp} are disregarded because they contribute to $D_2$ only at higher order in $\text{e}^{-1/(2\alpha^2)}$.

The expansions of $I_0,I_1,$ and $I_2$ along with Eqs.~\eqref{eq:Iexpand} and ~\eqref{eq:beta1} gives $D_2$ to the next-to-leading-order non-analytic term, Eq.~\eqref{eq:d2_1}. In this expansion of $D_2$, each non-analytic term is multiplied by an alternating, divergent series in $\alpha$. We use Pad\'e-Borel resummation with Pad\'e approximants of order $(32,32)$ to extract meaningful information from these series \cite{Gus16}.
\section{Mellin-Barnes transforms}\label{sec:BMT}
The asymptotic expansion of the function $_1F_1^{(1,0,0)}(1,1/2,1/{2 a^2})$ for $a \to 0$ can be calculated using its Mellin-Barnes representation. The Mellin-Barnes representation is well known in the fields of Finite Temperature Quantum Field Theory where it is used to find asymptotics of infinite sums, and Conformally invariant Quantum Field Theories where it is the natural substitute for the Fourier representation due to its scale invariance properties. The discussion in this introductory section closely follows \cite{Par01}.

The Mellin transform $F(s)$ of a function $f(x)$ is defined as
\begin{align} \label{mellin_def}
 F(s) \equiv \mathcal{M}[f;s]   = \int_0^\infty \mathrm {d}x \ x^{s-1} f(x),
\end{align}
and the transform can be inverted to give
\begin{align}
 f(x) = \int_\mathcal{C} \mathrm{d}s \ x^{-s} F(s),
\end{align}
where $\mathcal{C}$ is a contour in the complex $s$ plane. $\mathcal{C}$ is typically a line parallel to the $y-$axis, a curve asymptoting as $|s| \to \infty$ in the second and third quadrant and intersecting the $x-$axis at a finite value of $\text{Re}(s)$, or a combination of the two. Assuming for $f(x)$, with $\delta > 0$,
\begin{align}
 f(x)= \begin{cases}
      O(x^{-a-\delta}), & x\to 0+ \\
      O(x^{-b+\delta}), & x\to +\infty
   \end{cases},
\end{align}
the integral \eqref{mellin_def}  is absolutely convergent and $F(s)$ is an analytic function in the strip $a < \text{Re}(s) < b$, referred to as the \textit{strip of analyticity} of $F(s)$. Typically, the function $F(s)$ can be analytically continued outside this strip. It can been proved that if $f(x)$ has the asymptotic behavior (compare with eq. \eqref{eq:trans})
\begin{align} \label{eq:mellin_asymp}
 f(x)= \begin{cases}
      {\rm e}^{- b_1 x^{-\mu_1}}  \sum_{m} \sum_{l=0}^{N_1(m)} c_{lm} (\log x)^l x^{a_m} \,, x\to 0+ \\
      {\rm e}^{- b_2 x^{\mu_2}} \ \ \sum_{m} \sum_{l=0}^{N_2(m)} c_{lm}^\prime (\log x)^l x^{-b_m}, x\to \infty. 
   \end{cases}
\end{align}
$F(s)$ may be continued to, at worst, a meromorphic function outside it's \textit{strip of analyticity}, and the singular terms in the Laurent expansion of $F(s)$ near the poles, if any, can be determined solely in terms of the constants appearing in \eqref{eq:mellin_asymp}, the asymptotic expansion of $f(x)$\cite{Ble86}.

\subsection{Calculation of Asymptotics of \texorpdfstring{$_1F_1^{(1,0,0)}(1,1/2,1/{2 a^2})$}{1f1}}
For $\text{Arg}(-z) < \frac{\pi}{2}$ the Confluent Hypergeometric function has the Mellin representation
\begin{align}
 _1F_1(a,b,z) = \frac{\Gamma(b)}{\Gamma(a)}\frac{1}{2 \pi i}\int_{c-i\infty}^{c+i\infty} \frac{\Gamma(-s)\Gamma(s+a)}{\Gamma(s+b)}(-z)^s \ \mathrm{d}s ,
\end{align}
where the contour of integration is a line parallel to the $y-$axis, possibly with kinks to separate the poles of $\Gamma(-s)$ from those of $\Gamma(s+a)$. One can obtain an asymptotic expansion of the $_1F_1$ by shifting his contour to the left over the poles of $\Gamma(s+a)$ \cite{Par01}. Here we'll use the same procedure to obtain the asymptotic expansion of $_1F_1^{(1,0,0)}$. We use the Kummer transformation,
\begin{align}
  _1F_1(a,b,z) = e^z \, _1F_1(b-a,b,-z),
\end{align}
followed by differentiation in the first argument of $_1F_1$ to obtain
\begin{align}
&e^{-z} {{}_1F_1}^{(1,0,0)}(1,\nicefrac{1}{2},z)=  \frac{d}{da} {}_1F_1(b-a,b,-z) \Big|_{a=1,b=1/2} \nonumber\\
 &= \psi(-\nicefrac{1}{2}) {}_1F_1(-\nicefrac{1}{2},\nicefrac{1}{2},-z) \nonumber\\
 &+ \frac{1}{2} \frac{1}{2 \pi i} \int_{c-i\infty}^{c+i\infty} \frac{\Gamma(-s)\Gamma(s-1/2)}{\Gamma(s+1/2)} \psi(s-1/2)\, z^s \ \mathrm{d}s \\
 &= \psi(-\nicefrac{1}{2}) {}_1F_1(-\nicefrac{1}{2},\nicefrac{1}{2},-z) \nonumber\\
 &+ \frac{1}{2} \frac{1}{2 \pi i} \int_{c-i\infty}^{c+i\infty} \frac{\Gamma(-s)}{s-1/2} \psi(s-1/2) \, z^s \ \mathrm{d}s.
\end{align}
Since $\Gamma(s-\nicefrac{1}{2})$ has poles at at $s = \nicefrac{1}{2}, -i-\nicefrac{1}{2} ; i \in \mathbb{N}^0$ we obtain an asymptotic expansion by shifting the contour to the left over these poles, which gives%
\begin{align}
 &\frac{1}{2 \pi i} \int_{c-i\infty}^{c+i\infty} \frac{\Gamma(-s)}{s-1/2} \psi(s-1/2) \, z^s \ \mathrm{d}s \nonumber\\
 	&= 2 \sqrt{\pi} (\gamma - \psi(-\nicefrac{1}{2}))\sqrt{z} + 2 \sqrt{\pi} \sqrt{z} \log z \nonumber\\
	& + \frac{1}{2 \pi i} \int_{c-1-i\infty}^{c-1+i\infty} \frac{\Gamma(-s)}{s-1/2} \psi(s-1/2) \, z^s \ \mathrm{d}s \nonumber\\
 	&= 2 \sqrt{\pi} (\gamma - \psi(-\nicefrac{1}{2}))\sqrt{z} + 2 \sqrt{\pi} \sqrt{z} \log z \nonumber\\
	&+ \sum_{n=0}^\infty \frac{\Gamma(n+\nicefrac{1}{2})}{n+1} \frac{1}{\sqrt{z} z^n}.
\end{align}
This gives the full asymptotic expression for $I_1$,%
\begin{align}
 &I_1(\alpha) \sim \frac{\alpha^2}{2} \sum_0^\infty\frac{(2k-1)!!}{k+1}\alpha^{2k} \nonumber \\
 &+\alpha \frac{e^{-\nicefrac{1}{2 \alpha^2}}}{\sqrt{2\pi}} (2 (\gamma - 1) - \log{\frac{\alpha^2}{2}})\sum_1^\infty (-1)^k(2k-1)!!\alpha^{2k}
\end{align}

%\bibliography{biblio}

%merlin.mbs apsrev4-1.bst 2010-07-25 4.21a (PWD, AO, DPC) hacked
%Control: key (0)
%Control: author (8) initials jnrlst
%Control: editor formatted (1) identically to author
%Control: production of article title (-1) disabled
%Control: page (0) single
%Control: year (1) truncated
%Control: production of eprint (0) enabled
%
\end{document}